\documentstyle[11pt]{article}
\gdef\Feynmanlength{\setlength{\unitlength}{0.01pt}}  

\global\newcount\LINETYPE
\global\newcount\LINEDIRECTION
\global\newcount\LINECONFIGURATION
\newcommand{\LTYPE}{\LINETYPE}
\newcommand{\LDIR}{\LINEDIRECTION}

\global\LINETYPE=1  \global\LINEDIRECTION=0  \global\LINECONFIGURATION=0
\global\newcount\fermion    \fermion=1
\global\newcount\scalar     \scalar=2
\global\newcount\photon     \photon=3
\global\newcount\gluon      \gluon=4
\global\newcount\SPECIAL    \SPECIAL=5
\gdef\N{0}  \gdef\NE{1}  \gdef\E{2}   \gdef\SE{3}
       
\global\newcount\REG            \global\REG=0
\global\newcount\FLIPPED        \global\FLIPPED=1
\global\newcount\CURLY          \global\CURLY=2
\global\newcount\FLIPPEDCURLY   \global\FLIPPEDCURLY=3
\global\newcount\FLAT           \global\FLAT=4
\global\newcount\FLIPPEDFLAT    \global\FLIPPEDFLAT=5
\global\newcount\CENTRAL        \global\CENTRAL=6
\global\newcount\FLIPPEDCENTRAL \global\FLIPPEDCENTRAL=7
             
\global\newcount\SQUASHEDGLUON  \global\SQUASHEDGLUON=8

\newcount\adjx \adjx=0
\newcount\adjy \adjy=0
\global\newdimen\BIGPHOTONS     \BIGPHOTONS=0pt  
\global\newdimen\THICKPHOTONS     \THICKPHOTONS=0pt  
\global\newdimen\THICKPHOTONSWITCH    \THICKPHOTONSWITCH=0pt
\gdef\THICKPHOTONTEST{
\THICKPHOTONSWITCH=0pt
\ifdim\THICKPHOTONS=0pt \relax
  \else \ifnum\LTYPE=3
           \ifnum\LDIR=2 \THICKPHOTONSWITCH=1pt \fi 
           \ifnum\LDIR=6 \THICKPHOTONSWITCH=1pt \fi 
        \fi
\fi
}  
\gdef\THICKLINES{\thicklines  \THICKPHOTONS=1pt}

\global\newcount\phantomswitch   \global\phantomswitch=0
\global\newcount\stemlength   \global\stemlength=275   
\global\newcount\absstemlength        
\global\newcount\stemlengthx          
\global\newcount\stemlengthy          
\newdimen\FRONTSTEM  \FRONTSTEM=0pt   
\newdimen\BACKSTEM   \BACKSTEM=0pt    
\newdimen\EITHERSTEM \EITHERSTEM=0pt  
\global\newcount\arrowlength                
\global\newdimen\ATTIP   \global\ATTIP=0pt  
\global\newdimen\ATBASE  \global\ATBASE=1pt 
\global\newcount\unitboxnumber  
\global\newcount\unitboxnumberpo  
\global\newcount\particlelengthx  
\gdef\plengthx{\particlelengthx}
\global\newcount\particlelengthy  
\gdef\plengthy{\particlelengthy}
\global\newcount\boxlengthx  
\global\newcount\boxlengthy  
\global\newcount\particleadjustx  
\global\newcount\particleadjusty  
\global\newcount\particlelength   
\global\newcount\particlefrontx
\gdef\pfrontx{\particlefrontx}
\global\newcount\PFRONTx
\global\newcount\particlefronty
\gdef\pfronty{\particlefronty}
\global\newcount\PFRONTy
\global\newcount\particlebackx
\gdef\pbackx{\particlebackx}
\global\newcount\particlebacky
\gdef\pbacky{\particlebacky}
\global\newcount\particlemidx
\gdef\pmidx{\particlemidx}
\global\newcount\particlemidy
\gdef\pmidy{\particlemidy}
\global\newcount\seglength  \global\newcount\gaplength
\global\gaplength=850  
\global\seglength=1416  
\global\newcount\Xone    \global\newcount\Yone    
\global\newcount\Xtwo    \global\newcount\Ytwo    
\global\newcount\Xthree  \global\newcount\Ythree  
\global\newcount\Xfour   \global\newcount\Yfour   
\global\newcount\Xfive   \global\newcount\Yfive   
\global\newcount\Xsix    \global\newcount\Ysix    
\global\newcount\Xseven  \global\newcount\Yseven  
\global\newcount\Xeight  \global\newcount\Yeight  
\newsavebox{\lastline}  
\global\newcount\numlineparts   
\global\newcount\upperlineadjx  \upperlineadjx=0  
\global\newcount\upperlineadjy  \upperlineadjy=0  
\global\newcount\lowerlineadjx  \lowerlineadjx=0  
\global\newcount\lowerlineadjy  \lowerlineadjy=0  
\global\newcount\thirdlineadjx  \thirdlineadjx=0  
\global\newcount\thirdlineadjy  \thirdlineadjy=0  
\global\newcount\fourthlineadjx \fourthlineadjx=0  
\global\newcount\fourthlineadjy \fourthlineadjy=0  
\global\newcount\unitboxwidth   \unitboxwidth=1000
\global\newcount\unitboxheight  \unitboxheight=0  
\global\newcount\numupperunits  \numupperunits=8  
\global\newcount\numlowerunits  \numlowerunits=8  
\global\newcount\numthirdunits  \numthirdunits=8  
\global\newcount\numfourthunits \numfourthunits=8  
\global\newcount\fermioncount   \global\fermioncount=0
\global\newcount\scalarcount    \global\scalarcount=0
\global\newcount\photoncount    \global\photoncount=0
\global\newcount\gluoncount     \global\gluoncount=0
\global\newcount\SPECIALcount   \global\SPECIALcount=0
\global\newcount\vertexcount    \global\vertexcount=-1
\global\newcount\XDIR
\global\newcount\YDIR
\gdef\SETDIR{  
\ifcase\LDIR
     \global\XDIR=0  \global\YDIR=1   
\or  \global\XDIR=1  \global\YDIR=1   
\or  \global\XDIR=1  \global\YDIR=0   
\or  \global\XDIR=1  \global\YDIR=-1  
\or  \global\XDIR=0  \global\YDIR=-1  
\or  \global\XDIR=-1 \global\YDIR=-1  
\or  \global\XDIR=-1 \global\YDIR=0   
\or  \global\XDIR=-1 \global\YDIR=1   
\else\DIRECTERROR
\fi}  
\gdef\moduloeight#1{
\ifnum#1>7 \global\advance #1 by -8
\relax
\moduloeight#1
\relax
\else \relax
\fi}
\gdef\multroothalf#1{\global\multiply #1 by 7071 \global\divide #1 by 10000}
\gdef\negate#1{\global\multiply #1 by -1}

\gdef\slanttest(#1,#2){
\ifodd\LDIR
\multiply #1 by 7071  \divide #1 by 10000
\multiply #2 by 7071  \divide #2 by 10000
\fi
}
\gdef\gslanttest(#1,#2){
\ifodd\LDIR
\multroothalf#1
\multroothalf#2
\fi
}
\gdef\setplength{ 
\global\particlelengthx=\unitboxwidth
\global\particlelengthy=\unitboxheight
\global\multiply \particlelengthx by \unitboxnumber
\global\multiply \particlelengthy by \unitboxnumber
\global\advance \particlelengthx by \particleadjustx
\global\advance \particlelengthy by \particleadjusty
}
\gdef\boxlengthdefault{  
\global\boxlengthx=\plengthx
\global\boxlengthy=\plengthy
\ifnum\plengthx<0 \global\multiply\boxlengthx by -1 \fi
\ifnum\plengthy<0 \global\multiply\boxlengthy by -1 \fi
}
\gdef\rearcoords{  
\global\particlebacky=\particlefronty
\global\particlebackx=\particlefrontx
\global\advance \particlebackx by \particlelengthx
\global\advance \particlebacky by \particlelengthy
}
\gdef\midcoords{  
\global\particlemidy=\particlefronty
\global\particlemidx=\particlefrontx
\global\stemlengthx=\particlelengthx  
\global\stemlengthy=\particlelengthy
\global\divide\stemlengthx by 2
\global\divide\stemlengthy by 2
\global\advance \particlemidx by \stemlengthx
\global\advance \particlemidy by \stemlengthy
}
\gdef\setparticle{\setplength\rearcoords\midcoords\boxlengthdefault}  
\gdef\setcoords(#1,#2,#3)(#4,#5,#6)[#7,#8]{
\global\upperlineadjx=#1
\global\lowerlineadjx=#2
\global\thirdlineadjx=#3
\global\upperlineadjy=#4
\global\lowerlineadjy=#5
\global\thirdlineadjy=#6
\global\unitboxwidth=#7
\global\unitboxheight=#8
}
\gdef\drawoldpic#1(#2,#3){  
\global\particlefrontx=#2
\global\particlefronty=#3
\rearcoords
\midcoords
\put(#2,#3){\usebox{#1}}
}
\gdef\drawsavedline`#1' as #2[#3#4](#5,#6)[#7]{
\global\LINETYPE=#2
\global\LINEDIRECTION=#3
\global\LINECONFIGURATION=#4
\global\particlefrontx=#5
\global\particlefronty=#6
\global\unitboxnumber=#7
\selectcase
\rearcoords
\midcoords
\ifnum\phantomswitch=0 \drawas{#1}\fi
}

\gdef\drawas#1{
\global\savebox{#1}(\boxlengthx,\boxlengthy){
\setlength{\unitlength}{0.01pt}
\begin{picture}(\boxlengthx,\boxlengthy)
\multiput(\upperlineadjx,\upperlineadjy)(\unitboxwidth,\unitboxheight)
{\numupperunits}{\upperunitbox}
\ifnum\numlineparts > 1  
\multiput(\lowerlineadjx,\lowerlineadjy)(\unitboxwidth,\unitboxheight)
{\numlowerunits}{\lowerunitbox}
\fi
\ifnum\numlineparts > 2  
\multiput(\thirdlineadjx,\thirdlineadjy)(\unitboxwidth,\unitboxheight)
{\numthirdunits}{\thirdunitbox}
\fi
\ifnum\numlineparts > 3  
\multiput(\fourthlineadjx,\fourthlineadjy)(\unitboxwidth,\unitboxheight)
{\numfourthunits}{\lowerunitbox}
\fi
\end{picture} }
\global\PFRONTx=\pfrontx  \global\PFRONTy=\pfronty   
\SETFRONTSTEM
\THICKPHOTONTEST
\ifdim\THICKPHOTONSWITCH=1pt\global\advance\PFRONTy by 20  \fi
\put(\PFRONTx,\PFRONTy) {\usebox{#1}}   
\ifdim\THICKPHOTONSWITCH=1pt
\global\advance\PFRONTy by -40
\put(\PFRONTx,\PFRONTy) {\usebox{#1}}   
\global\advance \PFRONTy by 20  
\fi  
\SETBACKSTEM
\seglength=1416   \gaplength=850   
}
\gdef\drawandsaveline`#1' as #2[#3#4](#5,#6)[#7]{
\global\newsavebox{#1}
\drawsavedline`#1' as #2[#3#4](#5,#6)[#7]
}

\gdef\drawline#1[#2#3](#4,#5)[#6]{   
\drawsavedline`\lastline' as #1[#2#3](#4,#5)[#6]}

\gdef\TYPEERROR{\message{*** ERROR IN PARTICLE TYPE SELECTION ***}
\message{+++ Try with line type \fermion,\scalar,\photon,\gluon
(see manual) +++}\SETERR}
\gdef\DIRECTERROR{\SETERR\message{*** ERROR IN PARTICLE DIRECTION SELECTION
***}
\message{+++ Try again with direction N, NE, E, SE  etc. or see manual +++}}
\gdef\UNIMPERROR{\message{*** ERROR IN PARTICLE OPTIONS SELECTION ***}
\message{
+++ The requested options combination has not yet been implemented +++}\SETERR}
\gdef\SETERR{\gdef\upperunitbox{{\tiny Error}}  
\gdef\lowerunitbox{\relax}
\gdef\thirdunitbox{\relax}
}
\gdef\neglengthcheck{\ifnum\unitboxnumber < 1
\message{   *** ERROR:  PARTICLE OF NEGATIVE OR ZERO LENGTH REQUESTED. ***   }
\message{   ***         TAKING ABSOLUTE VALUE. ***   }\negate\unitboxnumber
\fi}
\gdef\selectcase{
\neglengthcheck   
\SETDIR
\ifcase\LINETYPE
\TYPEERROR  
\or \selectfermion  
\or \selectscalar   
\or \selectphoton   
\or \selectgluon    
\or \selectspecial  
\else \TYPEERROR \fi  }
\gdef\selectfermion{
\ifnum\fermioncount=0 
\global\newcount\fermionlength  
\global\newcount\fermionlengthx
\global\newcount\fermionlengthy
\global\newcount\fermionfrontx  
\global\newcount\fermionfronty  
\global\newcount\fermionbackx
\global\newcount\fermionbacky
\gdef\ALLfermion{  
\global\fermionfrontx=\particlefrontx \global\fermionfronty=\particlefronty
\ifnum\unitboxnumber > 50000
\message{   *** WARNING *** Fermion of length
\the\unitboxnumber\space requested ***   }
\ifnum\unitboxnumber > 80000
\message{   *** Reducing fermion length to 30000 (max 80000) ***   }
\global\unitboxnumber=30000 \fi \fi  
\global\fermionlength=\unitboxnumber 
\global\particleadjustx=0   \global\particleadjusty=0 
\global\numlineparts = 1    \global\numupperunits=1
\global\upperlineadjx=-200  \global\upperlineadjy=0
\global\fermionlengthx=\fermionlength    \global\fermionlengthy=\fermionlength
\gslanttest(\fermionlengthx,\fermionlengthy)  
\global\multiply\fermionlengthx by \XDIR  
\global\multiply\fermionlengthy by \YDIR  
\global\unitboxheight=\fermionlengthy   \global\unitboxwidth=\fermionlengthx
\global\advance \fermionlengthx by \particleadjustx
\global\advance \fermionlengthy by \particleadjusty
\global\particlelengthx=\fermionlengthx
\global\particlelengthy=\fermionlengthy
\boxlengthdefault    \rearcoords    \midcoords
\global\fermionbackx=\particlebackx     \global\fermionbacky=\particlebacky
\ifcase\LINECONFIGURATION  
\ifnum\XDIR=0
\gdef\upperunitbox{\line(\XDIR,\YDIR){\boxlengthy}} 
\else
\gdef\upperunitbox{\line(\XDIR,\YDIR){\boxlengthx}}
\fi
\else \UNIMPERROR
\fi
}

 \fi
\global\advance\fermioncount by 1  
\ALLfermion
}
\gdef\selectscalar{
\ifnum\scalarcount=0 
\newcount\scalarlength
\newcount\scalarlengthx
\newcount\scalarlengthy
\newcount\scalarfrontx  
\newcount\scalarfronty  
\newcount\scalarbackx
\newcount\scalarbacky
\gdef\ALLscalar{
\global\scalarfrontx=\particlefrontx   
\global\scalarfronty=\particlefronty   
\numlineparts = 1      \numupperunits=\unitboxnumber
\ifcase\LINECONFIGURATION
\global\upperlineadjx=-200     \global\upperlineadjy=0
\slanttest(\seglength,\gaplength)   
\gdef\upperunitbox{\line(\XDIR,\YDIR){\seglength}}
\else \UNIMPERROR 
\fi
\global\unitboxwidth=\seglength  \global\advance\unitboxwidth by \gaplength
\global\multiply \unitboxwidth by \XDIR
\global\unitboxheight=\seglength  \global\advance\unitboxheight by \gaplength
\global\multiply \unitboxheight by \YDIR
\global\particleadjustx=\gaplength \global\multiply\particleadjustx by \XDIR
\global\particleadjusty=\gaplength \global\multiply\particleadjusty by \YDIR
\negate\particleadjustx   \negate\particleadjusty   
\setparticle  
\global\scalarlengthx=\particlelengthx  
\global\scalarlengthy=\particlelengthy  
\ifnum\boxlengthx > 50000
\message{   *** WARNING *** Scalar of length in excess of 50000cp
requested!}\fi
\ifnum\boxlengthy > 50000
\message{   *** WARNING *** Scalar of length in excess of 50000cp
requested!}\fi
\global\scalarbackx=\pbackx      \global\scalarbacky=\pbacky   
}

 \fi
\global\advance\scalarcount by 1  
\ALLscalar
}
\gdef\selectphoton{   
\ifnum\photoncount=0 \input PHOTONSETUP  \fi
\selectphoton
}
\gdef\selectgluon{   
\ifnum\gluoncount=0 \input GLUONSETUP  \fi
\selectgluon
}
\gdef\selectspecial{\UNIMPERROR}
\gdef\checkvertex{ 
\ifnum\vertexcount=-1   \input VERTEX  \fi}
\gdef\drawvertex#1[#2#3](#4,#5)[#6]{\checkvertex\drawvertex#1[#2#3](#4,#5)[#6]}
\gdef\vertexcap#1{\checkvertex\vertexcap#1}
\gdef\vertexcaps{\checkvertex\vertexcaps}
\gdef\vertexlink#1{\checkvertex\vertexlink#1}
\gdef\vertexlinks{\checkvertex\vertexlinks}
\gdef\stemvertex#1{\checkvertex\stemvertex#1}
\gdef\stemvertices{\checkvertex\stemvertices}
\gdef\flipvertex{\checkvertex\flipvertex}
\global\arrowlength=349  
\gdef\drawarrow[#1#2](#3,#4){
\global\LDIR=#1
\SETDIR
\global\boxlengthx=#3  
\global\boxlengthy=#4  
\ifdim#2=1pt  
\adjx=\arrowlength      \adjy=\arrowlength
\multiply\adjx by \XDIR \multiply\adjy by \YDIR  
\slanttest(\adjx,\adjy)
\global\advance\boxlengthx by \adjx    \global\advance\boxlengthy by \adjy
\fi
\ifnum\phantomswitch=0\put(\boxlengthx,\boxlengthy){\vector(\XDIR,\YDIR){0}}\fi
}  
\gdef\SETFRONTSTEM{
\EITHERSTEM=\FRONTSTEM   \advance\EITHERSTEM by \BACKSTEM
\ifdim\EITHERSTEM>0pt
\global\stemlengthx=\stemlength   \global\stemlengthy=\stemlength
\global\absstemlength=\stemlength
\SETDIR
\gslanttest(\stemlengthx,\stemlengthy)
\gslanttest(\absstemlength,\REG)  
\ifnum\XDIR=0 \stemlengthx=0 \fi
\ifnum\YDIR=0 \stemlengthy=0 \fi
\global\multiply\stemlengthx by \XDIR
\global\multiply\stemlengthy by \YDIR
\ifdim\FRONTSTEM=1pt
\ifnum\phantomswitch=0
          \put(\pfrontx,\pfronty){\line(\XDIR,\YDIR){\absstemlength}}\fi
\global\advance\plengthx by \stemlengthx
\global\advance\plengthy by \stemlengthy
\global\advance\PFRONTx by \stemlengthx
\global\advance\PFRONTy by \stemlengthy
\global\advance\pmidx by \stemlengthx
\global\advance\pmidy by \stemlengthy
\global\advance\pbackx by \stemlengthx
\global\advance\pbacky by \stemlengthy
\ifnum\LTYPE=3
\global\photonfrontx=\PFRONTx  \global\photonfronty=\PFRONTy
\global\photonbackx=\pbackx    \global\photonbacky=\pbacky
\fi  
\ifnum\LTYPE=4
\global\gluonfrontx=\PFRONTx  \global\gluonfronty=\PFRONTy
\global\gluonbackx=\pbackx    \global\gluonbacky=\pbacky
\fi  
\fi  
\fi  
}    
\gdef\SETBACKSTEM{
\ifdim\BACKSTEM=1pt
\ifnum\phantomswitch=0
       \put(\pbackx,\pbacky){\line(\XDIR,\YDIR){\absstemlength}}\fi
\global\advance\plengthx by \stemlengthx
\global\advance\plengthy by \stemlengthy
\global\advance\pbackx by \stemlengthx
\global\advance\pbacky by \stemlengthy
\fi  
\global\stemlength=275  \FRONTSTEM=0pt  \BACKSTEM=0pt 
}    
\gdef\drawloop#1[#2#3](#4,#5){  
\input LOOPS  
\drawloop#1[#2#3](#4,#5)}
\Feynmanlength  

\makeatletter
\def\gsim{\compoundrel>\over\sim}
\def\lsim{\compoundrel<\over\sim}
\def\compoundrel#1\over#2{\mathpalette\compoundreL{{#1}\over{#2}}}
\def\compoundreL#1#2{\compoundREL#1#2}
\def\compoundREL#1#2\over#3{\mathrel
         {\vcenter{\hbox{$\m@th\buildrel{#1#2}\over{#1#3}$}}}}
\makeatother
\def\sbullet{\scriptscriptstyle\bullet}
\begin{document}
\title{The Fermion Mass Hierarchy and Neutrino Mixing Problem}
\author{ C. D. Froggatt \\ {\em Department of Physics and
Astronomy,} \\ {\em University of Glasgow,} \\ {\em Glasgow G12 8 QQ,} \\
{\em Scotland, U.K.} \\}
\date{Email address: c.froggatt@physics.gla.ac.uk}
\maketitle

\begin{abstract}
The fermion mass problem is briefly reviewed. The observed
hierarchy of quark and charged lepton masses strongly suggests the
existence of an approximately conserved chiral flavour
symmetry beyond the Standard Model. It is argued that in
models of this type, the requirement of a natural explanation
for both the atmospheric and solar neutrino problems leads
to an essentially unique picture of neutrino masses and
mixing angles. The anti-grand unification model is used as an
explicit example to illustrate these ideas.
\end{abstract}
\vspace{0.8cm}
\section{Introduction}
The most striking feature of the charged fermion spectrum is
the hierarchy of quark-lepton masses,
ranging from the top quark with a mass
of order the electroweak scale, $M_t = 175$ GeV, down to the
electron of mass $1/2$ MeV. It therefore seems that the
top quark mass may be understood in terms of physics
already present at the electroweak scale, i.e.\ the Standard
Model (SM) or possibly its minimal supersymmetric
extension (MSSM); whereas the suppression of the other
fermion masses requires flavour dynamics beyond the SM or MSSM.

One popular mechanism for generating the top quark mass is
to assume that, at some high energy scale $M_X$, the running
Yukawa coupling constant $g_t(M_X)$ for the top quark is of order
unity or larger. It is attracted to its infra-red
quasi-fixed point value and, in the case of the MSSM, leads
to a successful prediction:
\begin{equation}
M_t \simeq (200\ \mbox{GeV})\ \sin\beta
\end{equation}
for $1.5 \lsim \tan\beta \lsim 2.5$. It is also possible
to get a large $\tan\beta$ solution, when all the third
generation Yukawa couplings, $g_t$, $g_b$, $g_{\tau}$,
contribute significantly to the renormalisation group
equations. I discussed this fixed point scenario at the sixth
Lomonosov conference two years ago \cite{lom95}. The
top quark mass has also been calculated in the SM, using the
so-called Multiple Point Principle (MPP) \cite{seoul}
according to which there should be another vacuum
with essentially the same energy
density as the usual SM vacuum. This principle requires the
top quark and Higgs pole
masses ($M_t$, $M_H$) to lie on the SM vacuum stability curve.
Furthermore the vacuum expectation value (VEV) of the Higgs field in
the second vaccuum is expected to be of the same order of
magnitude as the SM cut-off scale, which we take to be the Planck mass,
giving the SM predictions:
\begin{equation}
M_t = 173 \pm 5 \ \mbox{GeV} \qquad M_H = 135 \pm 9 \ \mbox{GeV}
\end{equation}
as first reported at the previous Lomonosov conference \cite{lom95}.

It is natural \cite{fn1} to try to explain the suppression of the other
SM fermion masses in terms of selection rules due to approximate
conservation laws. The mass $m$ in the Dirac equation is essentially a
transition amplitude between a left-handed fermion component
$\psi_L$ and its right-handed partner $\psi_R$. If $\psi_L$
and $\psi_R$ have different quantum numbers under an
approximate chiral symmetry
group G, the mass term is suppressed. So we are led to consider
introducing a chiral flavour (gauge) symmetry beyond the SM group
which, when unbroken, allows only the top quark Yukawa coupling
to be non-zero. The other quark and lepton masses and mixing
angles are then generated at some order in the VEV(s) responsible
for breaking the symmetry, measured relative to the fundamental
mass scale of the theory. Previously \cite{lom95} I illustrated
this mechanism for generating the fermion mass hierarchy using models
which extend the MSSM by an Abelian $U(1)_f$ flavour
symmetry \cite{ross,ramond} with Green-Schwarz anomaly cancellation.
As pointed out in section 2, it is possible to generate a realistic
mass hierarchy using an anomaly free $SM \otimes U(1)^2$ model.
In this talk I will use the anti-grand unification model \cite{seoul}
(AGUT) to illustrate this approach to the charged fermion mass
hierarchy and also to the neutrino mixing problem.

\section{AGUT Model and Fermion Mass Hierarchy}

The AGUT model \cite{seoul} is based on extending the SM gauge group,
$SMG = S(U(2)\otimes U(3)) \approx SU(3) \otimes SU(2) \otimes U(1)$,
to the non-simple gauge group $SMG^3 \otimes U(1)_f$ near the
Planck scale $M_{Planck} \simeq 10^{19}$ GeV. This means there is
a pure SM desert, without supersymmetry, up to within an order of
magnitude or so below $M_{Planck}$, and the SM gauge coupling
constants are not unified but their values are predicted \cite{larisa}
using the MPP. This AGUT group $SMG^3 \times U(1)_f$ at first seems
rather complicated and arbitrary. In fact it can be rather simply
characterised, as the maximal anomaly free subgroup $G_{max}$ of
the group $U(45)$ of unitary transformations on the known quark
and lepton Weyl fields, for which the SM irreducible representations
remain irreducible under $G_{max}$. However the main motivation
for considering this group is provided by its successful
phenomenological predictions/fits. The SM group is embedded in
$G_{max}$ as the diagonal subgroup of $SMG^3$ and, above the AGUT
breaking energy scale, each of the three quark-lepton generations
has its own set of SM-like gauge particles together with an Abelian
$U(1)_f$ gauge boson. The $SMG^3$ quantum numbers are assigned to
the quarks and leptons in the obvious way and the $U(1)_f$ charges
$Q_f$ are carried by just the right-handed fermions of the second and
third proto-generations:
\begin{equation}
Q_f(\tau_R) = Q_f(b_R) = Q_f(c_R) = 1
\quad Q_f(\mu_R) = Q_f(d_R) = Q_f(t_R) = -1
\end{equation}

\begin{figure}
\begin{picture}(24000,11000)(-5600,0)
\THICKLINES

\drawline\fermion[\E\REG](0,1500)[6000]
\drawarrow[\E\ATBASE](\pmidx,\pmidy)
\global\advance \pmidy by -2000
\put(\pmidx,\pmidy){$b_L$}

\put(6000,0){$\lambda_1$}

\drawline\fermion[\E\REG](6000,1500)[6000]
\drawarrow[\E\ATBASE](\pmidx,\pmidy)
\global\advance \pmidy by -2000
\put(\pmidx,\pmidy){$M_F$}

\put(12000,0){$\lambda_2$}

\drawline\fermion[\E\REG](12000,1500)[6000]
\drawarrow[\E\ATBASE](\pmidx,\pmidy)
\global\advance \pmidy by -2000
\put(\pmidx,\pmidy){$M_F$}

\put(18000,0){$\lambda_3$}

\drawline\fermion[\E\REG](18000,1500)[6000]
\drawarrow[\E\ATBASE](\pmidx,\pmidy)
\global\advance \pmidy by -2000
\put(\pmidx,\pmidy){$b_R$}

\drawline\scalar[\N\REG](6000,1500)[4]
\global\advance \pmidx by 1000
\global\advance \pmidy by 1000
\put(\pmidx,\pmidy){$\Phi_{WS}$}
\global\advance \scalarbackx by -530
\global\advance \scalarbacky by -530
\drawline\fermion[\NE\REG](\scalarbackx,\scalarbacky)[1500]
\global\advance \scalarbacky by 1060
\drawline\fermion[\SE\REG](\scalarbackx,\scalarbacky)[1500]

\drawline\scalar[\N\REG](12000,1500)[4]
\global\advance \pmidx by 1000
\global\advance \pmidy by 1000
\put(\pmidx,\pmidy){$W$}
\global\advance \scalarbackx by -530
\global\advance \scalarbacky by -530
\drawline\fermion[\NE\REG](\scalarbackx,\scalarbacky)[1500]
\global\advance \scalarbacky by 1060
\drawline\fermion[\SE\REG](\scalarbackx,\scalarbacky)[1500]

\drawline\scalar[\N\REG](18000,1500)[4]
\global\advance \pmidx by 1000
\global\advance \pmidy by 1000
\put(\pmidx,\pmidy){$T$}
\global\advance \scalarbackx by -530
\global\advance \scalarbacky by -530
\drawline\fermion[\NE\REG](\scalarbackx,\scalarbacky)[1500]
\global\advance \scalarbacky by 1060
\drawline\fermion[\SE\REG](\scalarbackx,\scalarbacky)[1500]

\end{picture}
\caption{Tree diagram for bottom quark mass in the AGUT model.
The crosses indicate the couplings of
the Higgs fields to the vacuum and the fundamental Yukawa
couplings $\lambda_i$ are of order unity.}
\label{MbFull}
\end{figure}
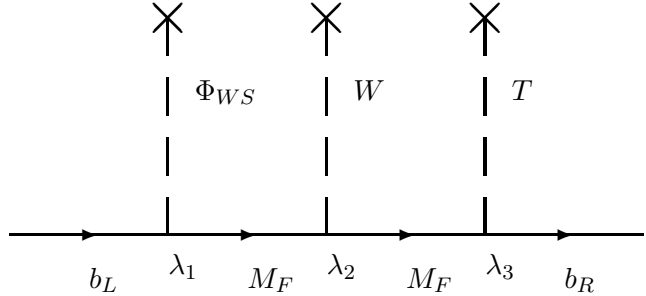

We now choose the Higgs fields responsible for the breakdown of the
$SMG^3 \otimes U(1)_f$ group to the SM group and the various
suppressions of the quark-lepton masses. Phenomenological
arguments lead us to introduce a Higgs field $S$ with a VEV of
order unity in Planck units and three Higgs fields $W$, $T$ and
$\xi$ with VEVs an order or magnitude smaller. Since the $S$-field
does not suppress the fermion mass matrix elements, phenomenological
arguments only determine the quantum numbers of the other Higgs
fields modulo those of $S$. With this choice of quantum numbers,
tree diagrams of the type shown in Fig.\ \ref{MbFull} generate
the following order of magnitude effective SM Yukawa coupling
matrices \cite{smg3m} for $u$ and $d$ type quarks:
\begin{equation}
\label{YUD}
Y_U \simeq \pmatrix{WT^2\xi^2 & WT^2\xi & W^2T\xi \cr
				   WT^2\xi^3 & WT^2    & W^2T    \cr
				   \xi^3     & 1       & WT \cr}
\qquad
Y_D \simeq \pmatrix{WT^2\xi^2 & WT^2\xi & T^3\xi  \cr
				   WT^2\xi   & WT^2    & T^3     \cr
				   W^2T^4\xi & W^2T^4  & WT	\cr}
\end{equation}
and
\begin{equation}
\label{YE}
Y_E \simeq \pmatrix{WT^2\xi^2 & WT^2\xi^3 & WT^4\xi \cr
				   WT^2\xi^5 & WT^2    & WT^4\xi^2 \cr
				   WT^5\xi^3 & W^2T^4  & WT \cr}
\end{equation}
for charged leptons.
Here $W$, $T$ and $\xi$ denote the VEVs of the Higgs field in
Planck units and we have assumed the presence of a rich spectrum
of vector-like Dirac fermions with fundamental masses
$M_F \simeq M_{Planck}$ to mediate the symmetry breaking transitions.
The corresponding set of Higgs field Abelian quantum numbers can be
specified as charge vectors $\vec{Q} \equiv \left ( y_1/2,
y_2/2,y_3/2,Q_f \right )$, where $y_i/2$ denotes the weak
hypercharge for the $i$'th proto-generation:
\begin{displaymath}
\vec{Q}_W = (-1/6 ,-1/3,1/2,-1/3) \quad
\vec{Q}_T = (-1/6,0,1/6 , 1/3) \quad
\vec{Q}_{\xi} = (0,0,0,1)
\end{displaymath}
\begin{equation}
\vec{Q}_S = ( 1/6,-1/6, 0 ,-1) \quad
\vec{Q}_{\Phi_{WS}} = (1/6 ,1/2 ,-1/6,0)
\end{equation}
The non-Abelian representations of the Higgs fields are,
like the fermions, taken to be singlet
or fundamental representations with their dualities and trialities
determined by the natural generalisation of the SM charge
quantisation rule.
The quantum numbers $\vec{Q}_{\Phi_{WS}}$ of the Weinberg-Salam
Higgs field $\Phi_{WS}$ are chosen to ensure
that the top quark Yukawa coupling
is of order one and corresponds to an off-diagonal element of $Y_U$.
We effectively only use the Abelian quantum numbers to determine
the mass suppression factors and, since we took $<S>=1$, we could
generate the same SM Yukawa matrix structure,
eqs.\ (\ref{YUD}) and (\ref{YE}),
with an anomaly free $SMG \otimes U(1)_{f1}\otimes U(1)_{f2}$ model
and a corresponding set of Higgs fields $W$, $T$, $\xi$ and
$\Phi_{WS}$. The two flavour charges
in such an Abelian extension of the SM could then be identified
as $Q_{f1} = y_3/2$ and $Q_{f2} = 4y_1/2 - 2y_2/2 + Q_f$.

The most characteristic feature of the AGUT Yukawa matrices $Y_U$,
$Y_D$ and $Y_E$ is that their diagonals are equal order of
magnitudewise.
Apart from the $t$ and $c$ quarks, the fermion mass eigenvalues
are given by the diagonal elements and
hence the AGUT model simulates the GUT SU(5) mass predictions,
namely the degeneracy of the $dsb$-quarks with the charged leptons
in the corresponding generations. However, we
only predict these degeneracies at the Planck scale
as far as order of magnitude is concerned, and not exactly!
This gives much better agreement with experiment than exact
SU(5) predictions, which are rather bad unless more
Weinberg-Salam Higgs fields are included
a la Georgi-Jarlskog's factor 3 mechanism \cite{georgijarlskog}.
Also note that we predict the $u$ quark to be
degenerate with the $d$ quark and the electron.
\begin{table}[t]
\caption{Best fit to experimental data. All masses are running masses at 1 GeV
except the top quark mass which is the pole mass.}
\begin{displaymath}
\begin{array}{|c|c|c|c|c|c|c|}
\hline
 & m_u & m_d & m_e & m_c & m_s & m_{\mu}  \\ \hline
{\rm Fit} & 3.6 {\rm \; MeV} & 7.0 {\rm \; MeV} &
0.87 {\rm \; MeV} & 1.02 {\rm \; GeV} & 400 {\rm \; MeV} &
88 {\rm \; MeV} \\ \hline
{\rm Data} & 4 {\rm \; MeV} & 9 {\rm \; MeV} &
0.5 {\rm \; MeV} & 1.4 {\rm \; GeV} & 200 {\rm \; MeV} &
105 {\rm \; MeV}  \\ \hline
\end{array}
\end{displaymath}
\begin{displaymath}
\begin{array}{|c|c|c|c|c|c|c|}
\hline
 & M_t & m_b & m_{\tau} & V_{us} & V_{cb} & V_{ub} \\ \hline
{\rm Fit}   & 192 {\rm \; GeV} & 8.3 {\rm \; GeV} &
1.27 {\rm \; GeV} & 0.18 & 0.018 & 0.0039 \\ \hline
{\rm Data}  & 180 {\rm \; GeV} & 6.3 {\rm \; GeV} &
1.78 {\rm \; GeV} & 0.22 & 0.041 & 0.0035 \\ \hline
\end{array}
\end{displaymath}
\label{bestfit}
\end{table}
In addition we have the following order
of magnitude Planck scale relations:
\begin{eqnarray}
m_b^3 \simeq m_t m_c m_s \qquad \qquad
V_{ub} \simeq  V_{td} \simeq V_{us} V_{cb}\\
V_{us} \simeq V_{cd} \simeq \sqrt{ \frac{ m_d}{m_s}} \qquad \qquad
V_{cb} \simeq V_{ts} \simeq \frac{ m_s^2}{ m_c m_b }
\end{eqnarray}
and predict the CP-violating area of the ``unitarity triangle''
to be given order of magnitudewise by $J \simeq V_{us} V_{cb} V_{ub}$.
The results of such a three parameter order of magnitude fit
to the data are given in Table 1.

\section{Neutrino Mixing Problem}

In this section we consider the generic textures of the neutrino mass
matrix, which arise in models with a natural fermion mass hierarchy
due to an approximately conserved chiral flavour symmetry.
We then require that the atmospheric and solar neutrino problems
be explained by the eigenvalues and mixing angles generated by
diagonalising the neutrino mass matrix. In this way we are led
to a picture \cite{fgn} in which the electron amd muon neutrinos are
quasi-degenerate in mass with maximal mixing between them, and both
atmospheric and solar neutrino data result from
$\nu_{\mu} \leftrightarrow \nu_e$ vacuum oscillations.

The effective three generation light neutrino mass matrix $M_{\nu}$,
generated by interactions beyond the SM, couples the left-handed
neutrinos with the right-handed anti-neutrinos:
\begin{equation}
{\cal L}_{m} = (M_{\nu})_{ij}\nu_{L_i}C\nu_{L_j} +\mbox{h.c.}
\label{Mnu}
\end{equation}
By its very definition $M_{\nu}$ is symmetric.
The overall neutrino mass scale is not really understand and
is usually set by hand, by an appropriate choice of $M_F$ in the
``see-saw" mass scale $\frac{\langle \Phi_{WS}\rangle^2}{M_{F}}$
or of the VEV of a weak isotriplet Higgs field $\Delta$. In models
with approximately conserved chiral charges, its matrix elements
are generally of different orders of magnitude, except for the
equalities enforced by the symmetry $M_{\nu} = M_{\nu}^T$.
The largest neutrino mass eigenvalue is then given by the
largest matrix element of $M_{\nu}$. If it
happens to be one of a pair of equal off-diagonal elements,
we get two very closely degenerate states as the heaviest
neutrinos and the third neutrino will be much lighter and,
in first approximation, will not mix with the other two \cite{fn2}.
If the largest element happens to be a diagonal element,
it will mean that the heaviest neutrino is a Majorana neutrino,
the mass of which is given by this matrix element, and it will
not be even order of magnitude-wise degenerate with the other,
lighter neutrinos. These lighter neutrinos may or may not get
their masses from off-diagonal elements and thus,
in first approximation, be degenerate.

The lepton mixing matrix $U$ is defined analogously to the usual
CKM quark mixing matrix, in terms of the unitary transformations
$U_{\nu}$ and $U_E$, on the left-handed lepton fields, which
diagonalise the squared neutrino mass matrix
$M_{\nu}M_{\nu}^{\dagger}$  and the squared charged lepton mass
matrix $M_EM_E^{\dagger}$ respectively:
\begin{equation}
U = U_{\nu}^{\dagger}U_E
\end{equation}
The charged lepton unitary transformation $U_E$ is expected to
be quasi-diagonal, with small off-diagonal elements due
to the charged lepton mass hierarchy. On the other hand when
there is a quasi-degenerate pair of neutrinos, because off-diagonal
elements dominate their masses, the
mixing angle contribution from $U_{\nu}$ will
be very close to $\pi/4$.

We are thereby led to consider the four possible textures given in
Table 2. With texture 1 the neutrino spectrum is hierarchical and
has small mixing angles like the charged fermion families.
Textures 2--4 correspond to having
a pair of quasi-degenerate neutrinos
with essentially maximal mixing and a third essentially unmixed
Majorana neutrino which may be $\nu_{\tau}$
or $\nu_{\mu}$ or $\nu_e$.

\begin{table}
\centering
\begin{tabular}{|c|c|c|c|} \hline
1 & $\pmatrix{A & \sbullet & \sbullet \cr
			   \sbullet & B & \sbullet \cr
			   \sbullet & \sbullet & C \cr}$  &
 Diagonal & \begin{tabular}{c}
 No strong mixings \\
 $ \theta$'s small
 \end{tabular}
 \\ \hline
2 & $\pmatrix{\sbullet & A & \sbullet \cr
			   A & \sbullet & \sbullet \cr
			   \sbullet & \sbullet & C \cr}$ &
\begin{tabular}{c}
$\nu_e \leftrightarrow \nu_{\mu}$ \\
mix strongly,\\
$\nu_{\tau}$ isolated
\end{tabular} &
\begin{tabular}{c}
$m_{\nu_e} \simeq m_{\nu_{\mu}}$ \\
$\sin^2 2\theta_{e\mu} \simeq 1$ \\
other $\theta$'s small
\end{tabular}
\\ \hline
3 & $\pmatrix{\sbullet & \sbullet & A \cr
			   \sbullet & B & \sbullet \cr
			   A & \sbullet & \sbullet \cr}$ &
\begin{tabular}{c}
$\nu_e \leftrightarrow \nu_{\tau}$ \\
mix strongly,\\
$\nu_{\mu}$ isolated
\end{tabular} &
\begin{tabular}{c}
$m_{\nu_e} \simeq m_{\nu_{\tau}}$ \\
$\sin^2 2\theta_{e\tau} \simeq 1$ \\
other $\theta$'s small
\end{tabular}
\\ \hline
4 & $\pmatrix{B & \sbullet & \sbullet \cr
			  \sbullet & \sbullet & A \cr
			   \sbullet & A & \sbullet  \cr}$ &
\begin{tabular}{c}
$\nu_{\mu} \leftrightarrow \nu_{\tau}$ \\
mix strongly,\\
$\nu_e$ isolated
\end{tabular} &
\begin{tabular}{c}
$m_{\nu_{\mu}} \simeq m_{\nu_{\tau}}$ \\
$\sin^2 2\theta_{\mu\tau} \simeq 1$ \\
other $\theta$'s small
\end{tabular}
\\ \hline
\end{tabular}
\caption{Neutrino mass matrix textures. The parameters $A$,
$B$ and $C$ are of different orders of magnitude, giving a hierarchy
of eigenvalues. The symbol $\sbullet$ is
used to denote relatively small elements responsible for small
mixings and small mass splittings between otherwise degenerate
eigenvalues. The mixing angles are estimated assuming the
contributions from the charged lepton matrix are small.}
\label{table:texture}
\end{table}

The atmospheric neutrino problem corresponds to a deficit of muon
neutrinos \cite{suzuki}, which could be explained by
$\nu_{\mu} \leftrightarrow \nu_{\tau}$ or
$\nu_{\mu} \leftrightarrow \nu_{e}$ oscillations
with $\Delta m^2_{atmos} \simeq 10^{-2}$ eV$^2$ and
strong mixing $\sin^2 2\theta \gsim 0.7$.
Textures 1 and 3 leave $\nu_{\mu}$ weakly mixed and are
thereby ruled out by our requirement of explaining the
atmospheric neutrino problem. We also want to explain the solar
neutrino problem and this requires mixing with the electron
neutrino. The only small mixing solution is the MSW solution
which has $\Delta m^2_{MSW} \simeq 10^{-5}$ eV$^2 \ll
\Delta m^2_{atmos}$. Thus the MSW solution would require
texture 4, but with a much greater degree of degeneracy
between $\nu_e$ and one of the other eigenstates than the
degeneracy between $\nu_{\mu}$ and $\nu_{\tau}$
naturally generated by the symmetry of the mass matrix. This
would require an extreme fine-tuning of parameters, which
we rule out as unnatural.

We are therefore left with a unique structure---texture 2---for
the neutrino mass matrix in our approach. This structure corresponds
to strong mixing of quasi-degenerate electron and muon neutrinos, with
an essentially isolated Majorana tau neutrino. Both the atmospheric
and solar neutrino problems are then solved by
$\nu_{\mu} \leftrightarrow \nu_e$ vacuum oscillations with close to
maximal mixing and $\Delta m^2_{e\mu} \simeq 10^{-2}$ eV$^2$. This
structure leads to an energy independent electron neutrino flux
suppression factor of 1/2 in all solar netrino experiments.
Also it requires the LSND evidence \cite{lsnd} for
$\bar{\nu_{\mu}} \rightarrow \bar{\nu_e}$ oscillations
should prove to be unfounded.
Since we have a herarchical structure, we expect the mass
splitting $\Delta m_{e\mu}$ to be at least one order of magnitude
smaller than $m_{\nu_e}$ and $m_{\nu_{\mu}}$. We can also use the
experimental limit \cite{suzuki} $m_{\nu_e} \lsim 10$ eV. So we
obtain the order of magnitude estimate
$m_{\nu_e} \simeq m_{\nu_{\mu}} \sim 1$ eV, which makes $\nu_e$
and $\nu_{\mu}$ candidates for hot dark matter. Due to the
hierarchical structure of $M_{\nu}$, the mass of the
tau neutrino should deviate from the other two
quasi-degenerate mass eigenvalues by
orders of magnitude. So, using the cosmological upper bound of
40 eV for stable neutrinos, we expect
$\nu_{\tau}$ to be much lighter than
$\nu_e$ and to be only slightly mixed.

It is possible to construct an explicit example \cite{fgn}
of such a neutrino mixing structure,
using the AGUT model with a triplet (under $SU(2)$
of the SM) Higgs field $\Delta$ having the Abelian charge vector:
\begin{equation}
\vec{Q}_W = (-1/2,-1/2,0,0)
\end{equation}
The corresponding charged lepton and neutrino mass matrices are:
\begin{equation}
M_E \sim \phi_{WS}  \pmatrix{
        WT^2\xi^2 & WT^2\xi^3 & WT^4\xi \cr
        WT^2\xi^5 & WT^2 & WT^4\xi^2 \cr
        WT^5\xi^3 & W^2T^4 & WT}
\quad
M_{\nu}  \sim  \Delta \pmatrix{
        \xi^3 & 1 & T^3\xi^2 \cr
        1 & \xi^3 & T^3 \xi \cr
        T^3\xi \rangle^2 & T^3\xi & T^3W^3\xi}
\end{equation}
which give, using the VEVs from the fit of Table 1,
$m_{\nu_e} \approx m_{\nu_{\mu}} \approx 2$ eV,
$m_{\nu_{\tau}} \approx 2 \times 10^{-7}$ eV,
$\Delta m^2_{e\mu} \approx 8 \times 10^{-3}$ eV$^2$
and $\sin^2 2\theta_{e\mu} \simeq 1$.

Our scenario of neutrino mixing is readily testable by
long baseline reactor neutrino oscillation experiments and, since
the Lomonosov conference, initial results from the CHOOZ
reactor experiment have become available \cite{CHOOZ97}.
This experiment finds (at 90\% confidence level) no evidence for
neutrino oscillations in the $\bar{\nu_e}$ disappearance mode for
$\Delta m^2 \gsim 10^{-3}$ eV$^2$ and maximum mixing, in
conflict with the $\nu_{\mu} \leftrightarrow \nu_e$ oscillation
solution to the atmospheric neutrino problem and, hence,
with our scenario. If both the atmospheric and solar neutrino
problems and the CHOOZ results are upheld, $M_{\nu}$
cannot\footnote{In principle the charged leptons could contribute
significantly to the $\mu-\tau$ mixing angle, via $U_E$,
without having the $\nu_{\mu}$ and $\nu_{\tau}$
masses quasi-degenerate. However
it is difficult to make the contribution from $U_E$ sufficiently
large that $\sin^2 2\theta_{\mu \tau} \gsim 0.7$
arises naturally in this way.}
have any of the textures given in Table 2 and we must conclude
that the dynamics underlying the structure of the neutrino mass
matrix is not understood.

\small

\end{document}